\documentclass[runningheads]{svmult}

\usepackage{makeidx}   
\usepackage{graphicx}  
\usepackage{subeqnar}  
\usepackage{multicol}  
\usepackage{physprbb}  
\makeindex             

\begin{document}

\title*{Percolation, Renormalization and the Quantum-Hall Transition}

\toctitle{Percolation, Renormalization and the Quantum-Hall Transition}

\titlerunning{Percolation and Quantum-Hall Transition}

\author{Rudolf A. R\"{o}mer}

\authorrunning{Rudolf A. R\"{o}mer}

\institute{Institut f\"{u}r Physik, Technische Universit\"{a}t,
           09107 Chemnitz, Germany
}

\maketitle

\begin{abstract}
  In this article, I give a pedagogical introduction and overview of
  percolation theory. Special emphasis will be put on the review of
  some of the most prominent of the algorithms that have been devised
  to study percolation numerically. At the central stage shall be the
  real-space renormalization group treatment of the percolation
  problem.  As a rather novel application of this approach to
  percolation, I will review recent results using similar real-space
  renormalization ideas that have been applied to the quantum Hall
  transition.
\end{abstract}

\section{Introduction}
\label{sec:intro}

Imagine a large chessboard\index{chessboard}, such as occasionally found
in a park.  It is fall, and all the master players have fled the cold a
long time ago. You are taking a walk and enjoy the beautiful sunny
afternoon, all the colors of the Indian summer\index{Indian
  summer} in the trees and in the falling leaves around you. Looking at
the chessboard, you see that some squares are already full of leaves,
while others are still empty \cite{fall}.  The pattern of the squares
which are covered by leaves seems rather random. As you try to cross the
chessboard, you see that there is a way to get from one side of the
board to the opposite side by walking on leaf-covered squares only. This
is percolation --- nearly.

Before continuing and explaining in detail what percolation is about,
let me outline the content of this paper. In Sect.\ 
\ref{sec:percolation}, I will review some of the most prominent and
interesting results on classical percolation. Percolation theory is at
the heart of many phenomena in statistical physics that are also
topics in this book. Beyond the exact solutions of percolation in
$d=1$ and $d=\infty$ dimensions, further exact solutions in $2\leq d
<\infty$ only rarely exist. Thus computational methods, using
high-performance computers and algorithms are needed for further
progress and in Sects.\ \ref{sec:color} -- \ref{sec:gradient}, I
explain in detail some of these algorithms. Section
\ref{sec:renormalization} is devoted to the real-space renormalization
group\index{real-space renormalization group} (RG)\index{RG} approach
to percolation.  This provides an independent and very suggestive
method of analytically computing results for the percolation problem
as well as a further numerical algorithm.

While many applications of percolation theory are mainly concerned with
problems of classical statistical physics, I will show in Sect.\
\ref{sec:qhe} that the percolation approach can give useful information
also at the quantum scale. In particular, I will show that aspects of
the quantum Hall (QH) effect can be understood by a suitably generalized
renormalization procedure of bond percolation in $d=2$. This application
allows the computation of critical exponents and conductance
distributions at the QH transition and also opens the way for studies of
scale-invariant, experimentally relevant macroscopic inhomogeneities. I
summarize in Sect.\ \ref{sec:concl}.

\section{Percolation}
\label{sec:percolation}

\subsection{The Physics of Connectivity}
\label{sec:connectivity}

From the chessboard example given above, we realize that the
percolation problem deals with the spatial {\em
  connectivity}\index{connectivity} of occupied squares instead of
simply counting whether the number of such squares has the majority of
all squares. Then the obvious question to ask is: How many leaves are
usually needed in order to allow passage across the board?  Since
leaves\index{leaves} normally do not interact with each other, and
friction-related forces can be assumed small compared to wind forces,
we can model the situation by assuming that the leaves are {\em
  randomly} distributed on the board. Then we can define an occupation
probability $p$ as being the probability that a site is occupied (by at
least one leaf\index{leaf}). Thus our question can be rephrased in
modern physics terminology as: Is there a threshold value $p_{\rm c}$ at
which there is a spanning cluster of occupied sites across an infinite
lattice?

The first time this question was asked and the term {\em percolation}
used was in the year 1957 in publications of Broadbent and Hammersley
\cite{BroH57,Ham57a,Ham57b}.  Since then a multitude of research
articles, reviews and books have appeared on this subject. Certainly
among the most readable such publications is the 1995 book by Stauffer
and Aharony \cite{StaA95}, where also most of the relevant research
articles have been cited. Let me here briefly summarize some of the
highlights that have been discovered in the nearly 50 years of research
on percolation.

The percolation problem in $d=1$ can be solved exactly. Since the number
of empty sites in a chain of length $L$ is $(1-p)L$, there is always a
finite probability for finding such an empty site in the infinite
cluster\index{infinite cluster} at $L\rightarrow \infty$ and thus the
percolation threshold is $p_{\rm c}=1$. Defining a correlation
function\index{correlation function} $g(r)\propto \exp(-r/\xi)$ which
measures the probability that a site at distance $r$ from an occupied
site at $0$ belongs to the same cluster, we easily find $g(r)=p^r$ and
thus $\xi
= -1/\ln p \approx (p_{\rm c}-p)^{-1}$. Thus close to the percolation
threshold\index{percolation threshold}, the correlation
length\index{correlation length} $\xi$ diverges with an exponent
$\nu=1$.

\begin{figure}[tbh]
\begin{center}
\includegraphics[width=0.9\textwidth]{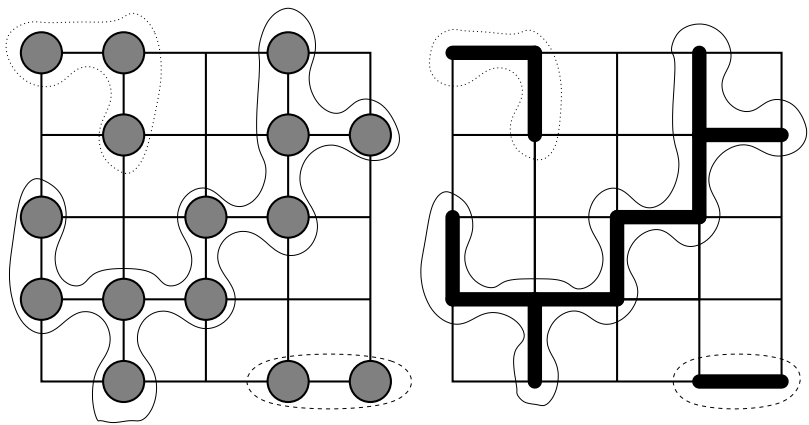}
\end{center}
\caption[]{
  {\em Site} (left) and {\em bond} (right) percolation on a square
  lattice. In site percolation, the sites of a lattice are occupied
  randomly and percolation is defined via, say, nearest-neighbor sites.
  In bond percolation, the bonds connecting the sites are used for
  percolation. The thin outlines define the $3$ clusters in each panel.
  The solid outline indicates the percolating cluster}
\label{fig:perco-site-bond}
\end{figure}

In $d=2$, the percolation problem provides perhaps the simplest example
of a second-order phase transition\index{second order phase
  transition}. The order parameter of this transition is the
probability $P(p)$ that an arbitrary site in the infinite lattice is
part of an infinite cluster\index{infinite cluster} \cite{infinite},
i.e.,
\begin{equation}
  \label{orderparam}
P(p) = \left\{
  \begin{array}{ll}
   0,               & p \leq {p_{\rm c}},\\ (p-{p_{\rm c}})^{\beta}, & p
   > {p_{\rm c}}
\end{array}\right.
\end{equation}
and $\beta$ is a critical exponent\index{critical exponent} similar to
the exponent $\nu$ of the correlation length. The distribution of the
sites in an infinite cluster\index{infinite cluster} at the
percolation threshold\index{percolation threshold} can be described as
a fractal\index{fractal} \cite{schreiber}, i.e., its average size $M$
in boxes of length $N$ increases as $\langle M(N) \rangle \propto
N^D$, where $D$ is the fractal dimension of the cluster. As in any
second-order phase transition, much insight can be gained by a
finite-size scaling\index{finite-size scaling} analysis
\cite{schreiber,Bin97}. In particular, the exponents introduced above
are related according to the scaling relation $\beta = (d-D)\nu$
\cite{StaA95}. Furthermore, it has been shown to an astonishing degree
of accuracy, that the values of the exponents and the relations
between them are independent of the type of lattice considered, i.e.,
square, triangular, honeycomb, etc., and also whether the percolation
problem is defined for sites\index{sites} or bonds\index{bonds} (see
Fig.\ \ref{fig:perco-site-bond}).  This independence is called {\em
  universality}. In the following, we will see that the
universality\index{universality} does not apply for the percolation
threshold\index{percolation threshold} $p_{\rm c}$. Thus it is of
importance to note that $p_{\rm c}$ for site percolation\index{site
  percolation} on the triangular lattice and bond percolation on the
square lattice is known exactly: $p_{\rm c}=1/2$. Especially the bond
percolation problem has received much attention also by mathematicians
\cite{Gri89}.

For higher dimensions, much of this picture remains unchanged, although
the values of $p_{\rm c}$ {\em and} the critical exponents change. The
upper critical dimension\index{upper critical dimension} corresponds to
$d=6$ such that mean field theory\index{mean field
  theory} is valid for $d\geq 6$ with exponents as given in Table
\ref{tab:critexp}.
\begin{table}[t]
\caption{\label{tab:critexp}
  Critical exponents $\nu$ and $\beta$ and fractal dimension $D$ for
  different spatial dimensions. For a more complete list see
  \cite{StaA95}.  }
\begin{center}
\renewcommand{\arraystretch}{1.4}
\setlength\tabcolsep{5pt}
\begin{tabular}[h]{lccccc}
\hline\noalign{\smallskip}
Exponent   & $d=2$   & $d=3$  & $d=4$  & $d=5$   & $d=6-\epsilon$ \\
\noalign{\smallskip}
\hline
\noalign{\smallskip}
${\nu}$    & $4/3$   & $0.88$ & $0.68$ & $0.57$  & $1/2 + 5\epsilon/84$ \\
$\beta$    & $5/36$  & $0.41$ & $0.64$ & $0.84$  & $1 - \epsilon/7$ \\

$D(p={p_{\rm c}})$ & $91/48$ & $2.53$ & $3.06$ & $3.54$  & $4 -
10\epsilon/21$\\ $D(p<{p_{\rm c}})$ & $1.56 $ & $2   $ & $12/5$ & $2.8 $
& $- $\\ $D(p>{p_{\rm c}})$ & $2    $ & $3   $ & $4   $ & $5   $  & $-
$\\
\hline
\end{tabular}
\end{center}
\end{table}

Applications of percolation theory are numerous \cite{BunH99}. It is
intimately connected to the theory of phase transitions\index{theory of
phase transitions} as discussed in \cite{Bin97,vojta}. The
connectivity\index{connectivity} problem is also relevant for diffusion
in disordered media \cite{schreiber,kramer} and networks \cite{grimm}. A
simple model of forest fires is based on percolation ideas
\cite{schwabl} and even models of stock market fluctuations \cite{Voi01}
have been devised using ideas of percolation \cite{GolLSJ00}.
Percolation is therefore a well-established field of statistical physics
and it continues its vital progress with more than 230 publications in
the cond-mat archives \cite{Con93} alone.

\subsection{The Coloring Algorithm}
\label{sec:color}

As stated above, there are only few exact results available in
percolation in $d\geq 2$. Thus in order to proceed further, one has to
use computational methods.

The standard numerical algorithm of percolation theory is due to Hoshen and
Kopelman \cite{HosK76}. Its advantage is due to the fact that it
allows to analyze which site belongs to which cluster without having
to store the complete lattice. Furthermore, this is being done in one
sweep across the lattice, thus reducing computer time.

At the heart of the Hoshen-Kopelman algorithm\index{Hoshen-Kopelman
  algorithm} is a bookkeeping mechanism \cite{StaA95}. Look at the site
percolation cluster in Fig.\ \ref{fig:perco-site-bond}. Going from left
to right and top to bottom through the cluster, we give each site a
label (or color) as shown in Fig.\ \ref{fig:hoshenkopelman}. If its top
or left neighbor is already occupied, then the new site belongs to the
same cluster and gets the same label. Otherwise we choose a new label.
In this way we can proceed through the cluster until we reach a problem
in line 3 as shown in the left column of Fig.\ \ref{fig:hoshenkopelman}.
According to our above rule the new site, indicated by the bold question
mark, can be either $2$ or $4$. This indicates that all sites previously
labelled by $2$ and $4$ actually belong to the same cluster. Thus we now
introduce an index\index{index of labels} Id$(\cdot)$ for each cluster
label and define it such that the index of a superfluous label, say $4$,
points to the right label, viz.\, Id$(4)=2$. Proceeding with our
analysis into the $4$th row of the lattice, we see in the center column
of Fig.\ \ref{fig:hoshenkopelman} that we again have to adjust our index
at the position indicated in bold.  Instead of labeling this site as
$4$, we choose $2\equiv$ Id$(4)$. And consequently, Id$(3)=2$. In this
way, we can easily check whether a cluster percolates from top to bottom
of the lattice by simply checking whether the label of any occupied site
in the bottom row of the lattice has an index equal to any of the labels
in the top row. Furthermore, in addition to the top row, we only need to
store the row presently under consideration and its predecessor. Thus
the storage requirement\index{storage requirement} is linear in lattice
size $L$ and not $L^2$ as it would be if we were to store the full
lattice. Last, the algorithm can also give information about all
clusters, say, if needed for a fractal analysis of the non-percolating
clusters. A Java implementation of such a coloring algorithm can be
found in Ref.\ \cite{Ada97}.

\begin{figure}[tbh]
\begin{center}
\renewcommand{\arraystretch}{1.4}
\setlength\tabcolsep{6pt}
{\large
\begin{tabular}{p{0.3\textwidth}p{0.3\textwidth}p{0.3\textwidth}}
\begin{tabular}[h]{|ccccc|}
\hline
$ 1        $&$ 1         $&$             $&$ 2         $&$ $ \\ \hline
$          $&$ 1         $&$             $&$ 2         $&$2 $  \\
$ 3        $&$           $&$ 4           $& {\bf ?}     &$ $  \\ \hline
$ {\bullet}$&$ {\bullet} $&$ {\bullet}   $&$           $&$ $  \\
$          $&$ {\bullet} $&$             $&$ {\bullet} $&$ {\bullet}$  \\
\hline
\end{tabular}
&
\begin{tabular}[h]{|ccccc|}
\hline
$ 1        $&$ 1         $&$             $&$ 2         $&$ $  \\ $ $&$ 1
$&$             $&$ 2         $&$2$  \\\hline $ 3 $&$           $&$ 4
$&$ 4         $&$ $  \\ $ 3        $&$ 3 $&  {\bf ?}     &$ $&$ $
\\\hline $          $&$ {\bullet} $&$             $&$ {\bullet}
$&${\bullet}$  \\
\hline
\end{tabular}
&
\begin{tabular}[h]{|ccccc|}
\hline
$ 1        $&$ 1         $&$             $&$ 2         $&$ $  \\
$          $&$ 1         $&$             $&$ 2         $&$2$  \\
$ 3        $&$           $&$ 4           $&$ 4         $&$ $  \\\hline
$ 3        $&$ 3         $&$      2      $&$           $&$ $  \\
$          $&$ { 2}      $&$             $&$ 5         $&$5$  \\
\hline
\end{tabular}
\\[1.5ex]
\renewcommand{\arraystretch}{1}
\setlength\tabcolsep{6pt}
\begin{tabular}[t]{lcr}
Id$({1})$& = &${1}$ \\
Id$({2})$& = &${2}$ \\
Id$({3})$& = &${3}$ \\
{\bf Id}{\boldmath $({4})$}& = &{\boldmath ${2}$}
\end{tabular}
&
\renewcommand{\arraystretch}{1}
\setlength\tabcolsep{6pt}
\begin{tabular}[t]{lcr}
Id$({1})$& = &${1}$ \\
Id$({2})$& = &${2}$ \\
{\bf Id}{\boldmath $({3})$}& = &{\boldmath ${2}$} \\
Id$({4})$& = &${2}$
\end{tabular}
&
\renewcommand{\arraystretch}{1}
\setlength\tabcolsep{6pt}
\begin{tabular}[t]{lcr}
Id$({1})$& = &${1}$ \\
Id$({2})$& = &${2}$ \\
Id$({3})$& = &${2}$ \\
Id$({4})$& = &${2}$ \\
Id$({5})$& = &${5}$
\end{tabular}
\end{tabular}
}
\end{center}
\caption{\label{fig:hoshenkopelman}
  Schematic description of the Hoshen-Kopelman coloring algorithm of the
  site percolation problem on a square lattice as shown in Fig.\
  \protect\ref{fig:perco-site-bond}. $\bullet$ denotes an occupied site,
  the numbers denote the cluster labels. The horizontal lines bracket
  the current and the previous row}
\end{figure}

\subsection{The Growth Algorithm}
\label{sec:growth}

\begin{figure}[tbh]
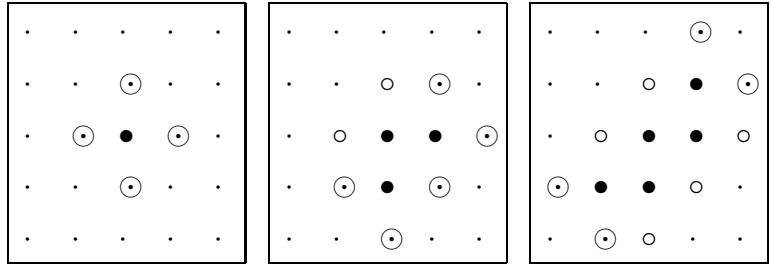

\begin{center}
\renewcommand{\arraystretch}{1.4}
\setlength\tabcolsep{6pt}
{\large
\begin{tabular}{p{0.25\textwidth}p{0.25\textwidth}p{0.25\textwidth}}
\begin{tabular}[h]{|p{6pt}p{6pt}p{6pt}p{6pt}p{6pt}|}
\hline
$\cdot    $&$ \cdot    $&$ \cdot   $&$ \cdot     $&$ \cdot   $ \\ $\cdot
$&$ \cdot    $&$ \odot   $&$ \cdot     $&$ \cdot   $ \\ $\cdot    $&$
\odot    $&$ {\bullet} $&$ \odot     $&$ \cdot   $ \\ $\cdot    $&$
\cdot    $&$ \odot   $&$ \cdot     $&$ \cdot   $ \\ $\cdot    $&$ \cdot
$&$ \cdot   $&$ \cdot     $&$ \cdot   $ \\
\hline
\end{tabular}
&
\begin{tabular}[h]{|p{6pt}p{6pt}p{6pt}p{6pt}p{6pt}|}
\hline
$\cdot    $&$ \cdot    $&$ \cdot   $&$ \cdot     $&$ \cdot   $ \\ $\cdot
$&$ \cdot    $&$ {\circ}   $&$ \odot     $&$ \cdot   $ \\ $\cdot    $&$
{\circ}  $&$ {\bullet} $&$ {\bullet}   $&$ \odot   $ \\ $\cdot    $&$
\odot    $&$ {\bullet} $&$ \odot     $&$ \cdot   $ \\ $\cdot    $&$
\cdot    $&$ \odot   $&$ \cdot     $&$ \cdot   $ \\
\hline
\end{tabular}
&
\begin{tabular}[h]{|p{6pt}p{6pt}p{6pt}p{6pt}p{6pt}|}
\hline
$\cdot    $&$ \cdot    $&$ \cdot   $&$ \odot     $&$ \cdot   $ \\ $\cdot
$&$ \cdot    $&$ {\circ}   $&$ \bullet  $&$ \odot  $ \\ $\cdot $&$
{\circ}    $&$ {\bullet} $&$ {\bullet}   $&$ \circ   $ \\ $\odot $&$
\bullet  $&$ {\bullet} $&$ \circ     $&$ \cdot   $ \\ $\cdot $&$
\odot    $&$ \circ   $&$ \cdot     $&$ \cdot   $ \\
\hline
\end{tabular}
\end{tabular}
}
\end{center}
\caption{\label{fig:leath}
  Schematic description of the first three steps in the Leath growth
  algorithm\index{growth algorithm} of the site percolation cluster of Fig.\
  \protect\ref{fig:perco-site-bond}. $\bullet$, $\circ$, and $\odot$
  denote an occupied, empty, and undecided site}
\end{figure}

When we want to study primarily the geometrical properties of
percolation clusters, another algorithm is more suitable which allows to
generate the desired cluster structure directly. This algorithm is due
to Leath \cite{Lea76} and works by a cluster-growth strategy. The idea
of the algorithm is that we put an occupied site in the center of an
otherwise empty lattice. Then we identify its nearest-neighbor sites as
shown in Fig.\ \ref{fig:leath}. Next we occupy these sites according to
the desired percolation probability\index{percolation probability} $p$.
We identify the new, yet undecided nearest-neighbor sites, occupy these
again with probability $p$ and repeat the procedure. The cluster
continues to grow until either all sites at the boundary are unoccupied
or the cluster has reached the boundary of the lattice. For $p<p_{\rm
c}$, the growth\index{growth} usually stops after a few iterations, while for
$p>p_{\rm c}$, percolating clusters are generated almost always. Thus
besides giving information about the fractal structure of the
percolating clusters, the Leath algorithm\index{Leath algorithm} can
also be used to estimate the value of $p_{\rm c}$. A Java implementation
of the Leath algorithm can be found in Refs.\ \cite{KinR98,KinR99}.

\subsection{The Frontier Algorithm}
\label{sec:gradient}

\begin{figure}[tbh]
\begin{center}
\includegraphics[width=0.58\textwidth]{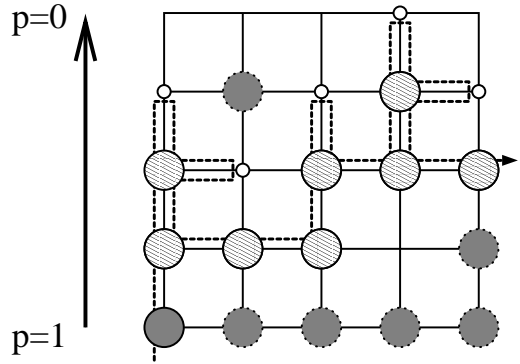}
\end{center}
\caption{
  An example for gradient percolation. All sites at $p=1$ ($p=0$) are
  occupied (empty). The $15$ large circles correspond to occupied
  sites.  The $8$ light shaded sites belong to the frontier, whereas
  the $7$ dark shaded sites are part of the interior of the cluster or
  belong to other clusters. The $5$ small circles correspond to sites
  in the empty frontier. The dashed line indicates the frontier
  generating walk. Note that only the $14$ sites with solid circles
  have actually been visited. The $6$ other circles are shown here
  just for clarity and need not be generated. According to
  (\protect\ref{eq:gradient-perco}), we have $p_{\rm c}= (8\cdot
  0.4375 + 5\cdot 0.75)/13=0.55769$}
\label{fig:gradient}
\end{figure}

As we have seen in the last section, the outer frontier\index{frontier}
of the percolation cluster is well defined. The fractal properties of
this hull can be measured \cite{Vos84,ZifCS84} and shown to yield
$D_{\rm h}=1.74\pm 0.02$. This suggests yet another algorithm for the
determination of $p_{\rm c}$ \cite{RosGS85,ZifS86}: Generate a lattice
with a constant gradient $\nabla p$ of the occupation probability $p$ as
shown in Fig.\ \ref{fig:gradient}.
By an algorithm similar to the one used in computing the hull of the
percolating cluster\index{hull of the percolating cluster}, one
traverses the frontier\index{frontier} of the occupied cluster and
determines which sites belong to it and which belong to the empty
cluster \cite{ZifCS84,RosGS85,ZifS86}. Then a very reliable estimate
for the percolation threshold can be computed as
\begin{equation}
  \label{eq:gradient-perco} p_{\rm c} = \frac{N_{\rm o} p_{\rm co} +
  N_{\rm e} p_{\rm ce}}{N_{\rm o} + N_{\rm e}},
\end{equation}
where $N_{o,e}$ denotes the number of sites in the occupied (empty)
frontier and $p_{co,ce}$ is the mean height of the associated
frontiers, respectively. Note that instead of actually generating the
percolation lattice, the algorithm instead proceeds by just generating
the sites needed for the construction of the frontier. Thus instead of
dealing with, say, ${\cal O}(L^2)$ sites as in the Hoshen-Kopelman and
Leath algorithms for a lattice in $d=2$, one only needs ${\cal O}(L)$
sites in the present algorithm \cite{frontier}. This reduces computer
time and estimates for $p_{\rm c}$ in a large variety of lattices can
be obtained with high accuracy
\cite{SudZ99,SykE63,ZifS97,LorZ98,KleZ98,NewZ00} as shown in Table
\ref{tab:pc}.
\begin{table}[t]
\caption{\label{tab:pc}
  Various current estimates of $p_{\rm c}$ for site and bond
  percolation on different lattices in $d=2$
  \cite{SudZ99,SykE63,ZifS97,LorZ98,KleZ98,NewZ00}.  Exactly known
  values are emphasized. The lattices are classified according to
  their number of first, second and more nearest-neighbors. The upper
  index gives the corresponding number in the dual lattice}
\begin{center}
\renewcommand{\arraystretch}{1.4}
\setlength\tabcolsep{5pt}
\begin{tabular}{llll}
\hline\noalign{\smallskip}
lattice &  & {site ${p_{\rm c}}$} & {bond ${p_{\rm c}}$}
\\
\noalign{\smallskip}
\hline
\noalign{\smallskip}
$3,12^{2}         $& &  {\em 0.807\,904}    & $$ \\
$4,6,12           $& &  {$0.747\,806$}    & $$ \\
$4,8^{2}          $& &  {$0.729\,724$}    & $$ \\

$6^{3}$  & honeycomb &  {$0.697\,043$}    & {\em 0.652\,703} \\

$3,6,3,6$ & Kagom\'e &  {\em 0.652\,703} & {$0.524\,4$} \\

$4^4$ & dice         &  {$0.584\,8$}        & {$0.475\,4$} \\

$3,4,6,4          $& &  {$0.621\,819$}    & $$ \\
$4^{4}$ & square     &  {$0.592\,746\,0$} & {\em 0.500\,000} \\
$3^{4},6          $& &  {$0.579\,498$}    & $$ \\
$3^{2},4,3,4      $& &  {$0.550\,806$}    & $$ \\
$3^{3},4^{2}      $& &  {$0.550\,213$}    & $$ \\

$3^{6}$ & triangular & {\em 0.500\,000}           & {\em 0.347\,296} \\
\hline
\end{tabular}
\end{center}
\end{table}

\section{Real-Space Renormalization}
\label{sec:renormalization}

\subsection{Making Use of Self-Similarity}
\label{sec:squinting}

As mentioned in Sect.\  \ref{sec:percolation}, the transition at $p_{\rm
c}$ corresponds to a second-order phase transition and the correlation
length $\xi$ is infinite. There is no particular length scale in the
system and all clusters are statistically similar to each other. This
{\em self-similarity}\index{self-similarity} \cite{schreiber} is at the
bottom of the renormalization\index{renormalization} description just as
for the fractal\index{fractal} analysis of the percolation clusters.

\begin{figure}[tbh]
\begin{center}
\includegraphics[width=0.5\textwidth]{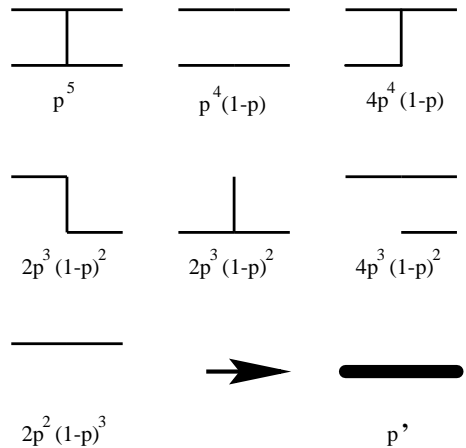}
\end{center}
\caption{
  The possible combinations of bonds (thin lines) that lead to a
  super-bond (thick line) together with their respective probabilities
  for bond percolation on a square lattice}
\label{fig:bond-rg}
\end{figure}

We may use the self-similarity\index{self-similarity} in the following
way. Let us replace a suitable collection of sites by {\em
  super}-sites and then study percolation of the super-lattice
\cite{HarLHD75,ReyKS77,Ber78,ReySK80,EscSH81}. In general, the
occupation probability\index{occupation probability} $p'$ of the
super-lattice will be different from the original $p$. Furthermore, if
the extent of the collection of sites in the lattice was $b$, then the
super-lattice will have a lattice constant $b$. Thus $\xi = b \xi'$ and
with $\xi \propto |p-p_{\rm c}|^{-\nu}$, we find
\begin{equation}
  \label{eq:self-sim} b |p'-p_{\rm c}|^{-\nu} \equiv |p-p_{\rm
  c}|^{-\nu}
\end{equation}
and consequently,
\begin{equation}
  \label{eq:nu-rg} \nu = \frac{\log b}{ \log\frac{\D p'}{\D
  p}}.
\end{equation}
As an example, let us consider the bond percolation\index{bond
  percolation} problem on a square lattice \cite{ReyKS77,Ber78,ReySK80}.
Here we replace 5 bonds by a super-bond\index{super-bond} in the
horizontal direction as shown in Fig.\ \ref{fig:bond-rg}.
Summing all probabilities for a connected, horizontal super-bond as
shown in Fig.\ \ref{fig:bond-rg}, we find that
\begin{equation}
  \label{eq:bond-rg}
   p'= p^5 + 5p^4(1-p) + 8 p^3(1-p)^2 + 2 p^2(1-p)^3.
\end{equation}
At the transition, we have $p'=p$ and thus (\ref{eq:bond-rg}) has the
solutions $p=0$, $0.5$, and $1$. The first and last solution correspond
to a completely empty or occupied lattice and are trivial. The second
solution reproduces the exact result of Table \ref{tab:pc}. From
\ref{eq:nu-rg}), we compute $\nu=1.4274$ which is already within $8\%$
of the exact result $4/3$. Thus the real-space RG\index{RG} scheme gives
very good approximations to the known results. But beware, it may not
always be that simple: the reader is encouraged to devise a similar RG
scheme for site percolation on a square lattice.

\subsection{Monte-Carlo RG}
\label{sec:monte-carlo}

\begin{figure}[tbh]
\begin{center}
\includegraphics[width=0.3\textwidth]{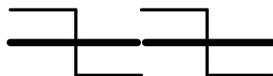}
\end{center}
\caption{
  Although the original bonds (thin lines) are not connected, the RG
  procedure outlined in the text nevertheless leads to two connected
  horizontal super-bonds (thick lines)}
\label{fig:bond-rg-wrong}
\end{figure}

The scheme of the last section is approximate since it cannot
correctly handle situations like the one in Fig.\ 
\ref{fig:bond-rg-wrong}.  In order to improve, we can construct an
RG\index{RG} scheme that uses a larger collection of bonds. The total
number of (connected and unconnected) configurations in such a
collection of $n$ bonds is $2^{n}$, putting severe bounds on the
practicability of the approach for analytic calculations. However, the
task is ideally suited for computers.  On the CD accompanying this
book, I include a set of {\sl Mathematica} routines that compute the
real-space RG for a $d=2$ triangular lattice.

\section{The Quantum-Hall Effect}
\label{sec:qhe}

In 1980, von Klitzing et al.\ \cite{KliDP80} found that the Hall
resistance\index{Hall resistance} $R_{\rm H}$ of MOSFETs at strong
magnetic field $B$ exhibits a step-like behavior which is accompanied
with a simultaneously vanishing longitudinal
resistance\index{longitudinal resistance} $R$. This is in contrast to
the classical Hall effect which gives a linear dependence of $R_{\rm H}$
on $B$.  Even more surprising, the values of $R_{\rm H}$ at the
transitions are given by universal constants, i.e., $\frac{1}{i}
\frac{h}{e^2}$, where $i$ is an integer.

Since its discovery this so-called integer quantum Hall
effect\index{integer quantum Hall effect} (IQHE)\index{IQHE} was
studied extensively \cite{JanVFH94,ChaP95}. Besides
semi-phenomenological models simply assuming a
localization-delocalization transition more general theories
considered, e.g., gauge invariance \cite{Lau81}, topological
quantization \cite{ThoKNN82}, scattering \cite{Pra81} and field
theoretical approaches \cite{Pru84}.

\subsection{Basics of the IQHE}
\label{sec:qhebasics}

\begin{figure}[tbh]
\begin{center}
\includegraphics[width=0.4\textwidth]{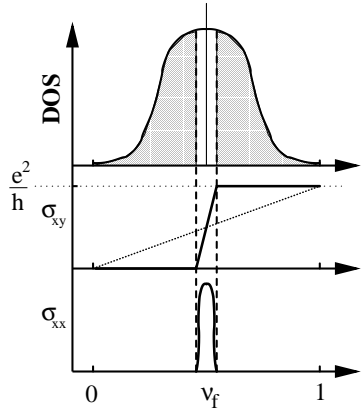}
\end{center}
\caption{
  Density of states (DOS), transversal and longitudinal conductivity
  as a function of $E_{\rm F}$ or, equivalently, filling factor
  $\nu_{\rm f}$ or $B^{-1}$ \cite{ChaP95}.  The peak in the middle of
  the band represents one $\delta$-function peak of the clean Landau
  model.  Dark shaded regions of the density of states correspond to
  localized states.  The thin dashed line (with non-zero slope) for
  $\sigma_{xy}$ indicates the classical Hall result}
\label{fig:qhe-basics}
\end{figure}

A simple understanding of the IQHE can be gained by considering the
Hamiltonian of a single electron in a magnetic field\index{magnetic
  field},
\begin{equation}
  H_0 = \frac{1}{2 m} \left( {\bf p} + \frac{e}{c} {\bf A} \right)^2
      = \frac{\hbar\omega_{\rm c}}{2 l_{B}^2} \left( \xi^2 + \eta^2
  \right). \label{eq:hamil}
\end{equation}
where $\vec{A}$ denotes the vector potential\index{vector potential} and
the Hamiltonian has been rewritten in guiding center coordinates
$X=x-\zeta$, $Y=y-\eta$ and relative coordinates $\zeta$, $\eta$
\cite{Lan30}.  Here, $\omega_{\rm c} =\frac{e B}{m}$ is the frequency of
the classical cyclotron motion and $l_{B} = (\frac{\hbar}{e B})^{1/2}$
is the radius of the cyclotron motion.  The spectrum of this Hamiltonian
is simply the harmonic oscillator with $E_n= \left( n +
  \frac{1}{2}\right) \hbar \omega_{\rm c},$ $n=0$, $1$, $\ldots$.  These
Landau levels\index{Landau levels} are infinitely degenerate since the
Hamiltonian no longer contains $X$ and $Y$. Thus the spectrum consists
of $\delta$-function peaks as indicated in Fig.\ \ref{fig:qhe-basics}.
Introducing disorder\index{disorder} into the model by adding a smooth
random potential $V(\vec{r})$ in (\ref{eq:hamil}) results in drift
motion\index{drift motion} of the guiding center\index{guiding center}
\begin{equation}
\dot{X}
  =\frac{\I}{\hbar}\left[H,X\right]  = \frac{l_{B}^2}{\hbar}
  \frac{\partial V}{\partial y} , \quad
\dot{Y}
  =\frac{\I}{\hbar}\left[H,Y\right]  = -\frac{l_{B}^2}{\hbar}
  \frac{\partial V}{\partial x}.
\end{equation}
perpendicular to the gradient of $V(\vec{r})$ (see Fig.\
\ref{fig:qhe-equi}). Furthermore, the degeneracy of the Landau levels is
lifted, the $\delta$-function density of states broadens
\cite{JanVFH94}, giving rise to a band-like structure as shown in Fig.\
\ref{fig:qhe-basics}. If the sample is penetrated by a {\em strong}
magnetic field, the cyclotron motion\index{cyclotron
  motion} is much smaller than the potential fluctuations. Consequently,
the electron motion can be separated into cyclotron motion and motion of
the guiding center along equipotential lines\index{equipotential lines}
of the energy landscape.

The IQHE can then be understood as follows: assume that the center of
the broadened Landau levels contain extended states that can support
transport, whereas the other states are spatially
localized\index{localized} and cannot. This is similar to the standard
picture in the theory of Anderson localization \cite{schreiber,kramer}.
Changing the Fermi energy\index{Fermi energy} $E_{\rm F}$ or the filling
factor\index{filling factor} $\nu_{\rm f} = 2 \pi l_{B}^2 \rho_{\rm e}= 2 \pi
\hbar \rho_{\rm e} / e B \propto E_{\rm F}$, where $\rho_{\rm e}$ denotes
the electron density, we first have $E_{\rm F}$ in the region of
localized states and both $\sigma_{xx}$ and $\sigma_{xy}$ are $0$. When
$E_{\rm F}$ reaches the region of extended states, there is transport,
$\sigma_{xx}$ is finite and $\sigma_{xy}=e^2/h$. Next, $E_{\rm F}$ again
reaches a region of localized states and $\sigma_{xx}$ drops back to $0$
until we reach the extended states\index{extended states} in the next
Landau level\index{Landau level}.

This picture suggests the following effective {\em classical} high-field
model\index{high field model} \cite{Ior82} of the IQHE: Neglecting the
cyclotron motion (i.e., large $B$) and quantum effects (i.e., only one
extended state) the classical electron transport with energy $E_{\rm F}$
through the sample only depends on the \lq\lq height" of the saddle
points in the potential energy landscape $V(\vec{r})$.  One obtains a
classical bond-percolation\index{bond percolation} problem
\cite{StaA95}, in which saddle points are mapped onto bonds. A
bond\index{bond} is connecting only when the potential of the
corresponding saddle point equals the energy of the electron $E_{\rm
F}$.  From percolation theory follows \cite{StaA95} that an infinite
system is conducting only when $E_{\rm F}=\langle V \rangle$.  Using
this model one could already describe the localization-delocalization
transition and thus the quantized plateaus in resistivity observed in
IQHE \cite{JanVFH94}.  But for bond percolation the correlation length
diverges at the transition with an exponent of $\nu=4/3$ which is in
contrast to the value found in the QH experiments.

\subsection{RG for the Chalker-Coddington Network Model}
\label{sec:rgcc}

\begin{figure}[tbh]
\begin{center}
  \begin{tabbing}
\includegraphics[width=0.6\textwidth]{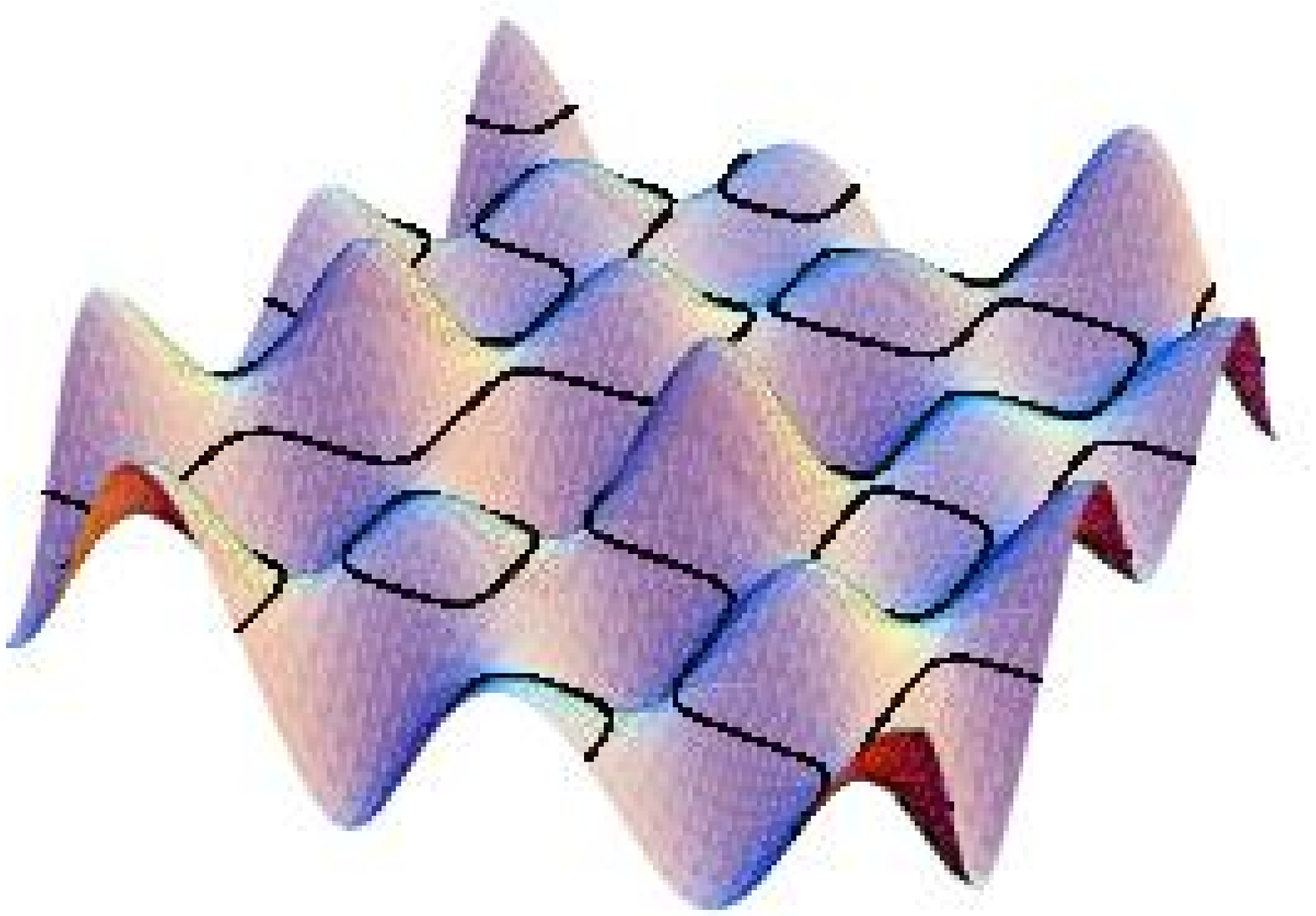}
\=
\includegraphics[width=0.4\textwidth]{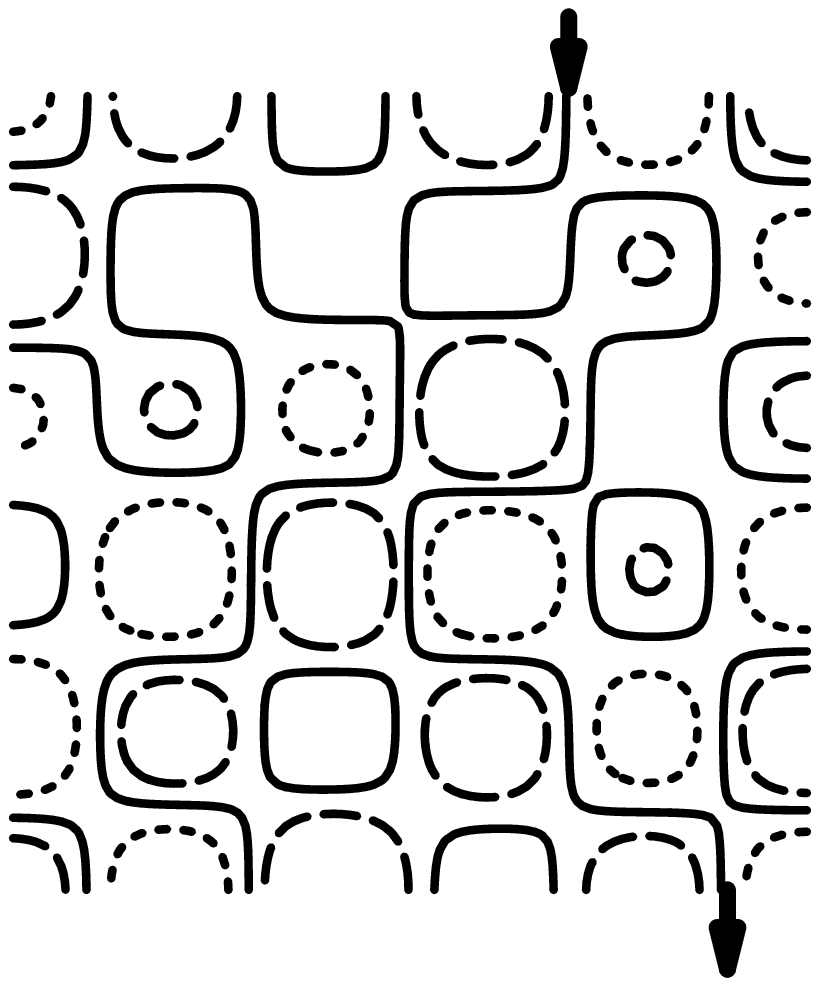}
  \end{tabbing}
\end{center}
\vspace*{-3ex}
\caption{
  Left: Schematic plot of a smooth random potential $V(\vec{r})$ with
  equipotential lines at $E=\langle V\rangle$ indicated in black.
  Right: Equipotential lines of the same potential for $E=\langle
  V\rangle-E_{\rm max}/2$, $\langle V\rangle$, and $\langle
  V\rangle+E_{\rm max}/2$ corresponding to long dashed, solid and
  short dashed lines. Note the solid line percolating the system
  from top to bottom as indicated by the arrows}
\label{fig:qhe-equi}
\end{figure}

The Chalker-Coddington (CC)\index{CC} network
model\index{Chalker-Coddington network model}\index{CC network model}
improved the high-field model by introducing quantum corrections
\cite{ChaC88}, namely tunneling\index{tunneling} and
interference\index{interference}.  Tunneling occurs, in a semiclassical
view, when electron orbits come close enough to each other and the
electron cyclotron motions overlap.  This happens at the saddle points,
which now act as quantum scatterers connecting two incoming with two
outgoing channels by a scattering matrix as shown in Fig.\
\ref{fig:qhe-equi}.
Similar to bond percolation a network can be constructed such that the
saddle points are mapped onto bonds.  While moving along an
equipotential line an electron accumulates a random phase which reflects
the disorder of $V(\vec{r})$. Results for this quantum
percolation\index{quantum percolation} also show one extended state in
the middle of the Landau band. The critical properties at the
transition, especially the value of the exponent $\nu\approx2.4$
\cite{LeeWK93}, agree with experiments \cite{KocHKP91,SchVOW00}.
\begin{figure}[tbh]
\begin{center}
  \begin{tabular}[t]{lr}
\includegraphics[width=0.2\textwidth]{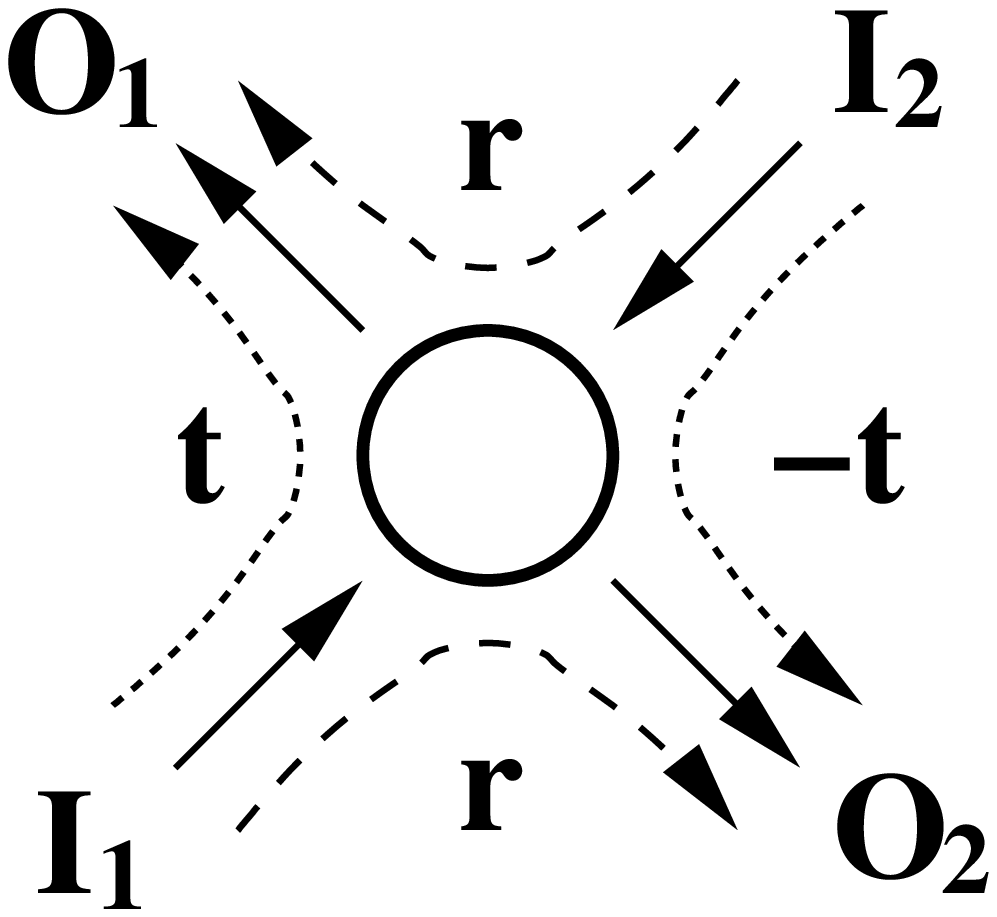}
\quad \quad
&
\quad \quad
\includegraphics[width=0.35\textwidth]{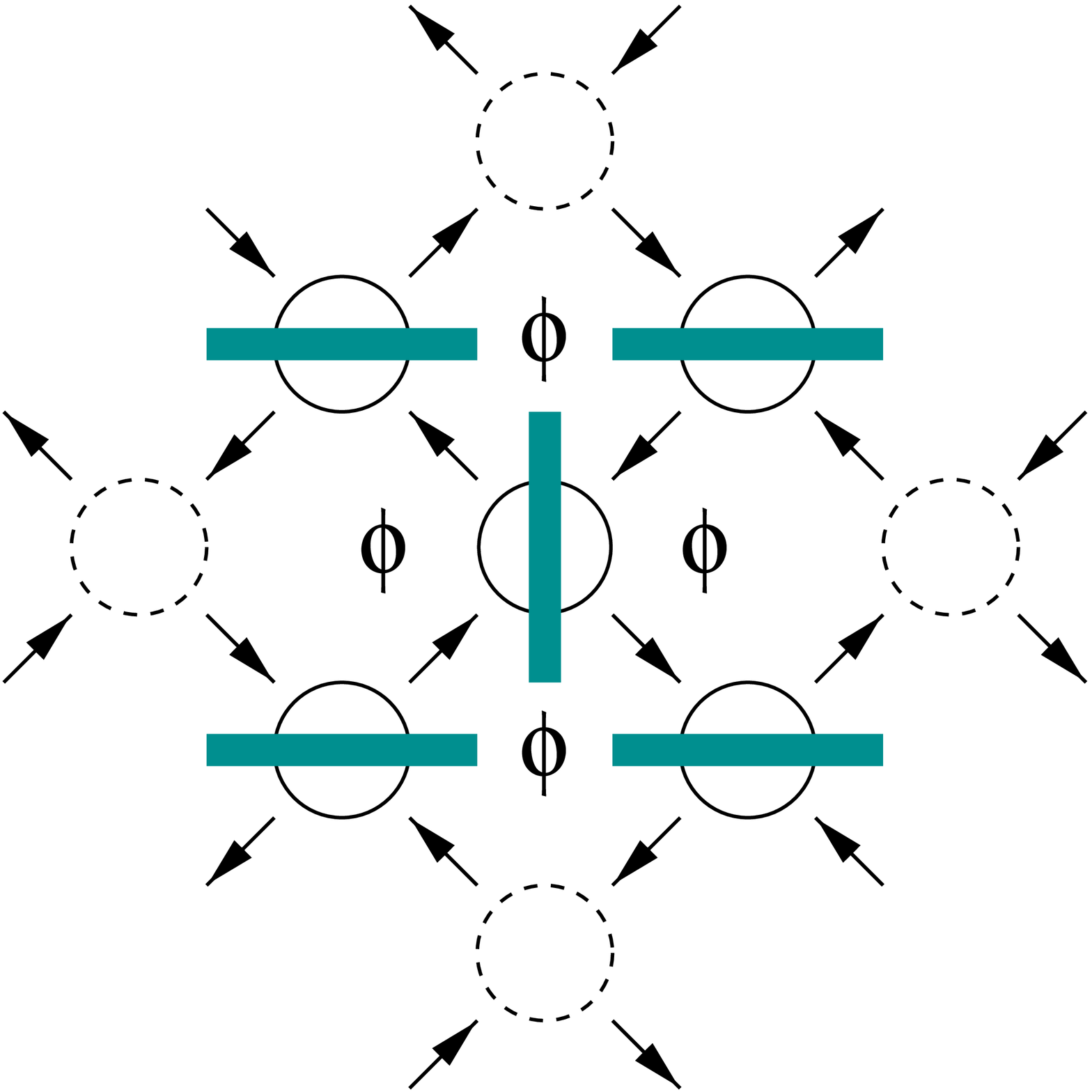}
  \end{tabular}
\end{center}
\caption{
  Left: A single saddle point (circle) connected to incoming and
  outgoing currents $I_i$, $O_i$ via transmission and reflection
  amplitudes $t$ and $r$.
  Right: A network of $5$ saddle points can be renormalized into a
  single super-saddle point by an RG approach very similar to the bond
  percolation problem of Sect.\  \protect\ref{sec:renormalization}. The
  phases are schematically denoted by the $\phi$'s}
\label{fig:rg-struct}
\end{figure}

As explained for the bond percolation problem we now apply the
RG\index{RG} method to the CC model. The RG structure which builds the
new super-saddle points is displayed in Fig.\ \ref{fig:rg-struct}. It
consists of $5$ saddle points drawn as bonds. The links (and phase
factors) connecting the saddle points are indicated by arrows pointing
in the direction of the electron motion due to the magnetic field.
Each saddle point acts as a scatterer connecting the $2$ incoming
$I_{1,2}$ with the $2$ outgoing channels $O_{1,2}$
\begin{equation}
\left(\begin{array}{c}
O_1\\
O_2
\end{array}
\right)
=
\left(
\begin{array}{cc}
t_i & r_i\\ r_i & -t_i
\end{array}
\right)
\left(
\begin{array}{c}
I_1\\
I_2
\end{array}
\right)
\end{equation}
with reflection coefficients\index{reflection coefficients} $r_i$ and
transmission coefficients\index{transmission coefficients} $t_i$, which
are assumed to be real numbers. The complex phase factors enter later
via the links between the saddle points.  By this definition --
including the minus sign -- the unitarity\index{unitarity} constraint
$t_i^2+r_i^2=1$ is fulfilled a priori. The amplitude of transmission of
the incoming electron to another equipotential line and the amplitude of
reflection and thus staying on the same equipotential line add up to
unity -- electrons do not get lost.

In order to obtain the scattering equation of the super-saddle point we
now need to connect the $5$ scattering equations according to Fig.\
\ref{fig:rg-struct}.  For each link the amplitude of the incoming
channels is defined by the amplitude of the outgoing channel of the
previous saddle point multiplied by the corresponding complex phase
factor $\E^{\I\phi_k}$.  This results in a system of $5$ matrix
equations, which has to be solved. One obtains an RG equation\index{RG
  equation} for the transmission coefficient $t'$ of the super-saddle
point \cite{GalR97} analogously to Eq.\ (\ref{eq:bond-rg}),
\begin{equation}
  \label{eq-qhrg} t'= \frac{ t_{15} (r_{234} \E^{\I\phi_2} - 1) + t_{24}
\E^{\I(\phi_{3}+\phi_{4})} (r_{135} \E^{-\I\phi_1} - 1) + t_3 (t_{25}
\E^{\I\phi_3} + t_{14} \E^{\I\phi_4}) } {
  (r_3 - r_{24} \E^{\I\phi_2}) (r_3 - r_{15} \E^{\I \phi_1}) + (t_3 -
t_{45} \E^{\I\phi_4}) (t_3 - t_{12} \E^{\I\phi_3}) }
\end{equation}
depending on the products $t_{i\ldots j}=t_i\cdot \ldots \cdot t_j$,
$r_{i\ldots j}=r_i\cdot \ldots\cdot r_j$ of transmission and
reflection coefficients $t_i$ and $r_i$ of the $i=1, \ldots, 5$ saddle
points and the $4$ random phases $\phi_k$ accumulated along
equipotentials in the original lattice. For further algebraic
simplification one can apply a useful transformation of the amplitudes
$t_i=(\E^{z_i}+1)^{-1/2}$ and $r_i = (\E^{-z_i}+1)^{-1/2}$ to heights
$z_i$ relative to heights $V_i$ of the saddle points. The
conductance\index{conductance} $G$ is connected to the transmission
coefficient $t$ by $G=|t|^2 e^2/h$ \cite{ButILP85}.

\subsection{Conductance Distributions at the QH Transition}
\label{sec:conductance}

For the numerical determination of the conductance distribution, we
first choose an initial probability distribution $P_0$ of transmission
coefficients $t$. The distribution is discretized in at least $1000$
bins. Thus the bin width\index{bin width} is typically $0.001
e/\sqrt{h}$ for the interval $t \in [0, e/\sqrt{h}]$.

Using the initial distribution $P_0(t)$, we now randomly select many
different transmission coefficients and insert them into the RG
equation\index{RG equation} (\ref{eq-qhrg}).  Furthermore, the phases
$\phi_j$, $j= 1,
\ldots 4$ are also chosen randomly, but according to a uniform
distribution $\phi_j \in [0, 2\pi]$. By this method at least $10^{7}$
super-transmission coefficients $t'$ are calculated and their
distribution $P_1(t')$ is stored. Next, $P_1$ is averaged using a
Savitzky-Golay smoothing\index{Savitzky-Golay smoothing} filter
\cite{PreFTV92} in order to decrease statistical fluctuations.  This
process is then repeated using $P_1$ as the new initial distribution.

\begin{figure}[tbh]
\begin{center}
    \includegraphics[width=0.72\textwidth]{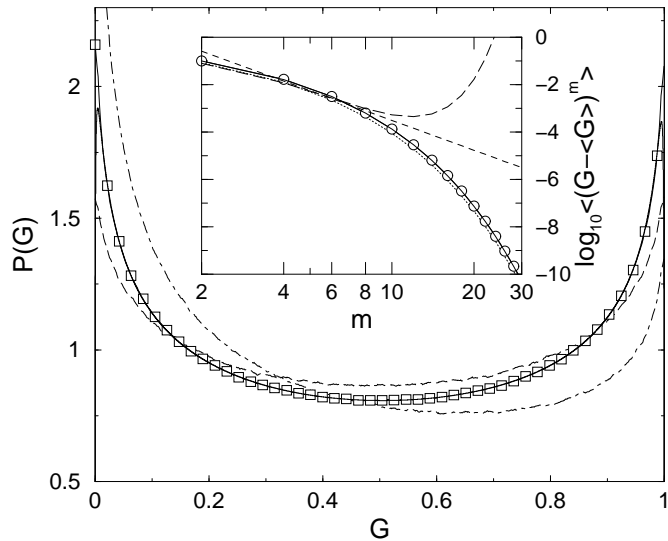}
\end{center}
\caption{
  Conductance distribution at a QH plateau-to-plateau transition. The
  squares correspond to the fixed-point distribution, dashed and
  dot-dashed lines to the initial distribution and an unstable
  distribution, respectively. The solid line indicates a fit of the FP
  distribution $P_{\rm c}(t)$ by three Gaussians.
  Inset: Moments of the FP distribution $P_{\rm c}(G)$. The dashed lines
  indicate various predictions based on extrapolations of results for
  small $m$ \cite{WanJL96}. The dotted line denotes the moments of a
  constant distribution}
\label{fig:rg-conductance}
\end{figure}
The iteration process is stopped when the distribution $P_i$ is no
longer distinguishable from its predecessor $P_{i-1}$ and we have
reached the desired fixed-point\index{fixed-point} (FP)\index{FP}
distribution\index{fixed-point distribution}\index{FP distribution}
$P_{\rm c}(t)$.  However, due to numerical instabilities, small
deviations from symmetry add up such that typically after $15$--$20$
iterations the distributions become unstable and converge towards the
classical FPs of no transmission or complete transmission similar to the
classical percolation case.  Figure \ref{fig:rg-conductance} shows this
behavior for one of the RG iterations\index{RG iterations}.  The FP
distribution $P_{\rm c}(G)$ shows a flat minimum around $G=0.5 e^2/h$
and sharp peaks at $G=0$ and $G=e^2/h$.  It is symmetric with $\langle G
\rangle = 0.498 e^2/h$.  This is in agreement with previous theoretical
\cite{WeyJ98,AviBB99} and experimental \cite{CobK96} results whereas our
results contain much less statistical fluctuations. Furthermore we
determine moments $\langle (G-\langle G\rangle)^m\rangle$ of the FP
distribution $P_{\rm c}(G)$. As shown in Fig.\ \ref{fig:rg-conductance}
for small moments up to $m=6$ our results agree with the work of Wang et
al.\ \cite{WanJL96}, who computed moments $m\leq 8.5$. But more
interesting is the fact that the obtained moments of the FP
distribution\index{moments of the FP
  distribution} can hardly be distinguished from the moments of a simple
constant distribution thus indicating the influence of the broad flat
minimum of the FP distribution around $G=0.5 e^2/h$.

\begin{figure}[tbh]
\begin{center}
\includegraphics[width=0.72\textwidth]{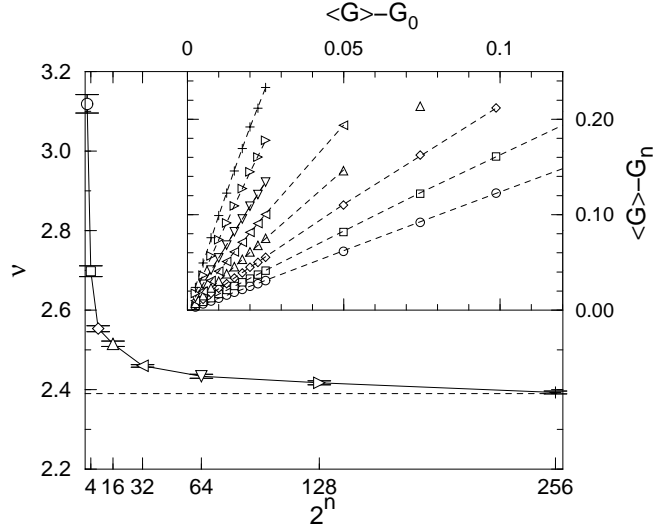}
\end{center}
\caption{
  Critical exponent $\nu$ as a function of magnification factor $2^n$
  for RG step $n$. The dashed line shows the expected result $\nu=2.39$.
  Inset: The shift of the average $G_n$ of $P(G)$ is linear in $G_0$.
  The dashed lines indicate linear fits to the data}
\label{fig:rg-nu-L}
\end{figure}

For the determination of the critical exponent\index{critical
  exponent}, we next perturb the FP distribution slightly, i.e., we
construct a distribution with shifted average $G_0$. Then we perform an
RG iteration\index{RG iteration} and compute the new average $G_1$ of
$P_1(G)$.  Tracing the shift of the perturbed average $G_n$ for several
initial shifts $G_0$, we expect to find a linear dependence of $G_n$ on
$G_0$ for each iteration step $n$.  The critical exponent is then
related to the slope $\D G_n/\D G_0$ \cite{CaiRRS00}.  Figure
\ref{fig:rg-nu-L} shows the resulting $\nu$ in dependence on the
iteration step and thus system size.  The curve converges close to
$\nu\approx 2.4$, i.e.\ the value obtained by Lee et al.\
\cite{LeeWK93}. Note that the \lq \lq system size" is more properly
called a system magnification, since we start the RG iteration with an
FP distribution valid for an infinite system and then magnify the system
in the course of the iteration by a factor $2^n$.

\section{Summary and Conclusions}
\label{sec:concl}

The percolation model represents the perhaps simplest example of a
system exhibiting complex behavior although its constituents -- the
sites and bonds -- are chosen completely uncorrelated. Of course, the
complexity enters through the connectivity requirement for percolating
clusters. I have reviewed several numerical algorithms for
quantitatively measuring various aspects of the percolation problem. The
specific choice reflects purely my personal preferences and I am happy
to note that other algorithms such as breadth- and depth-first
algorithms \cite{AhoHU83} have been introduced by P.\ Grassberger in his
contribution \cite{grassberger}.

The real-space RG\index{RG} provides an instructive use of the
underlying self-similarity\index{self-similarity} of the percolation
model at the transition.  Furthermore, it can be used to study very
large effective system sizes. This is needed in many applications. As
an example, I briefly reviewed and studied the QH transition and
computed conductance distributions, moments and the critical exponent.
These results can be compared to experimental measurements and shown
to be in quite good agreement.

\section*{Acknowledgements}
The author thanks Phillip Cain, Ralf Hambach, Mikhail E.\ Raikh, and
Andreas R\"{o}sler for many helpful discussions. This work was
supported by the NSF-DAAD collaborative research grant INT-9815194, the
DFG within SFB~393 and the DFG-Schwerpunktprogramm
``Quanten-Hall-Systeme''.


\begin{thebibliography}{50.}
\addcontentsline{toc}{chapter}{References}
\newcommand{\enquote}[1]{``#1''}
\expandafter\ifx\csname url\endcsname\relax
  \def\url#1{{\tt #1}}\fi
\expandafter\ifx\csname urlprefix\endcsname\relax\def\urlprefix{URL }\fi

\bibitem{fall} We note that the WEH-Ferienkurs, following which this article
  has been prepared, took place during {\em early} fall.

\bibitem{BroH57}
S.R. Broadbent, J.M. Hammersley: Proc. Camb. Philos. Soc. {\bf 53\/}, 629
  (1957)

\bibitem{Ham57a}
J.M. Hammersley: Proc. Camb. Philos. Soc. {\bf 53\/}, 642 (1957)

\bibitem{Ham57b}
J.M. Hammersley: Ann. Math. Statist. {\bf 28\/}, 790 (1957)

\bibitem{StaA95}
D.~Stauffer, A.~Aharony: {\em Perkolationstheorie\/} (VCH, Weinheim 1995)

\bibitem{infinite} For any finite lattice, \lq \lq infinite" means a
  cluster that reaches from top to bottom and/or left to right through
  the lattice.

\bibitem{schreiber} M. Schreiber, F. Milde (this volume).

\bibitem{Bin97}
K.~Binder: Rep. Prog. Phys. {\bf 60\/}, 487 (1997)

\bibitem{Gri89}
G.~Grimmett: {\em Percolation\/} (Springer, Berlin 1989)

\bibitem{BunH99}
A.~Bunde, S.~Havlin (Eds.): {\em Percolation and Disordered Systems: Theory and
  Applications\/} (North-Holland, Amsterdam 1999)

\bibitem{vojta} T. Vojta, (this volume).
\bibitem{kramer} B. Kramer, (this volume).
\bibitem{grimm} U. Grimm, (this volume).
\bibitem{schwabl} K. Schenk, B. Drossel, F. Schwabl, (this volume).
\bibitem{Voi01}
J.~Voit: {\em The Statistical Mechanics of Capital Markets\/} (Springer,
  Heidelberg 2001)

\bibitem{GolLSJ00} J.~Goldenberg, B.~Libai, S.~Solomon, N.~Jan,
  D.~Stauffer: {\em Marketing Percolation} (2000).  Cond-mat/9905426

\bibitem{Con93}
{\tt http://de.arXiv.org/}, 1993--2000

\bibitem{HosK76}
J.~Hoshen, R.~Kopelman: Phys. Rev. B {\bf 14\/}, 3438 (1976)

\bibitem{Ada97}
C.~Adami:  (1997), \\
{\tt
  http://www.krl.caltech.edu/\verb+~+adami/CD1/Percolation/percolation.html},
  likely to change without prior notice

\bibitem{Lea76}
P.~Leath: Phys. Rev. B {\bf 14\/}, 5056 (1976)

\bibitem{KinR98}
W.~Kinzel, G.~Reents: {\em Physics by Computer\/} (Springer, Berlin 1998)

\bibitem{KinR99}
W.~Kinzel, G.~Reents:  (1999),\\
 {\tt
  http://wptx15.physik.uni-wuerzburg.de/TP3/applet\verb+_+java/percgr.html},
  likely to change without prior notice

\bibitem{Vos84}
R.F. Voss: J. Phys. A: Math. Gen. {\bf 17\/}(7),
  L373 (1984)

\bibitem{ZifCS84}
R.M. Ziff, P.T. Cummings, G.~Stell: J. Phys. A: Math. Gen. {\bf 17\/},
  3009 (1984)

\bibitem{RosGS85}
M.~Rosso, J.F. Gouyet, B.~Sapoval: Phys. Rev. B {\bf 32\/}, 6053 (1985)

\bibitem{ZifS86}
R.M. Ziff, B.~Sapoval: J. Phys. A: Math. Gen. {\bf 19\/}, L1169 (1986)

\bibitem{frontier} Alas, this intuitively convincing argument is not
  strictly true: The percolation frontier is a fractal and as
  such scales $\propto L^{1.75}$ \cite{Vos84}. On the other hand, it
  is not the random number generation for $L^2$ sites in the
  Hoshen-Kopelman algorithm but rather the numerical determination of
  the percolating clusters which is numerically challenging.

\bibitem{SudZ99}
P.N. Suding, R.M. Ziff: Phys. Rev. E {\bf 60\/}, 275 (1999)

\bibitem{SykE63}
M.F. Sykes, J.W. Essam: Phys. Rev. Lett. {\bf 10\/}, 3 (1963)

\bibitem{ZifS97}
R.M. Ziff, P.N. Suding: J. Phys. A: Math. Gen. {\bf 30\/}, 5351 (1997)

\bibitem{LorZ98}
C.D. Lorenz, R.M. Ziff: Phys. Rev. B {\bf 57\/}, 230 (1998)

\bibitem{KleZ98}
P.~Kleban, R.M. Ziff: Phys. Rev. B {\bf 57\/}, R8075 (1998)

\bibitem{NewZ00}
M.E.J. Newman, R.M. Ziff:   (2000). Cond-mat/0005264

\bibitem{HarLHD75}
A.B. Harris, T.C. Lubensky, W.K. Holcomb, C.~Dasgupta: Phys. Rev. Lett. {\bf
  35\/}, 327 (1975)

\bibitem{ReyKS77}
P.J. Reynolds, W.~Klein, H.E. Stanley: J. Phys. C: Solid State Phys. {\bf
  10\/}, L167 (1977)

\bibitem{Ber78}
J.~Bernasconi: Phys. Rev. B {\bf 18\/}, 2185 (1978)

\bibitem{ReySK80}
P.J. Reynolds, H.E. Stanley, W.~Klein: Phys. Rev. B {\bf 21\/}, 1223
  (1980)

\bibitem{EscSH81}
P.D. Eschbach, D.~Stauffer, H.~Herrmann: Phys. Rev. B {\bf 23\/}, 422
  (1981)

\bibitem{KliDP80}
K.v. Klitzing, G.~Dorda, M.~Pepper: Phys. Rev. Lett. {\bf 45\/}, 494
  (1980)

\bibitem{JanVFH94}
M.~Janssen, O.~Viehweger, U.~Fastenrath, J.~Hajdu: {\em Introduction to the
  {Theory} of the {Integer} {Quantum} Hall effect\/} (VCH, Weinheim 1994)

\bibitem{ChaP95}
T.~Chakraborty, P.~{Pietil\"{a}nen}: {\em The Quantum {Hall} effects\/}
  (Springer, Berlin 1995)

\bibitem{Lau81}
R.B. Laughlin: Phys. Rev. B {\bf 23\/}, 5632 (1981)

\bibitem{ThoKNN82}
D.J. Thouless, M.~Kohmoto, M.P. Nightingale, M.~den Nijs: Phys. Rev. Lett. {\bf
  49\/}, 405 (1982)

\bibitem{Pra81}
R.E. Prange: Phys. Rev. B {\bf 23\/}, 4802 (1981)

\bibitem{Pru84}
A.M.M. Pruisken: Nucl. Phys. B {\bf 235\/}, 277 (1984)

\bibitem{Lan30}
D.L. Landau: Z. Phys. {\bf 64\/}, 629 (1930)

\bibitem{Ior82}
S.V. Iordanskii: Solid State Commun. {\bf 43\/}, 1 (1982)

\bibitem{ChaC88}
J.T. Chalker, P.D. Coddington: J. Phys.: Condens. Matter {\bf 21\/}, 2665
  (1988)

\bibitem{LeeWK93}
D.H. Lee, Z.~Wang, S.~Kivelson: Phys. Rev. Lett. {\bf 70\/}, 4130 (1993)

\bibitem{KocHKP91}
S.~Koch, R.J. Haug, K.~v.~Klitzing, K.~Ploog: Phys. Rev. B {\bf 43\/},
  6828 (1991)

\bibitem{SchVOW00}
R.T.F. van Schaijk, A.~de~Visser, S.M. Olsthoorn, H.P. Wei, A.M.M. Pruisken:
  Phys. Rev. Lett. {\bf 84\/}, 1567 (2000)

\bibitem{GalR97}
A.G. Galstyan, M.E. Raikh: Phys. Rev. B {\bf 56\/}, 1422 (1997)

\bibitem{ButILP85}
M.~{B\"{u}ttiker}, Y.~Imry, R.~Landauer, S.~Pinhas: Phys. Rev. B {\bf 31\/},
  6207 (1985)

\bibitem{PreFTV92}
W.H. Press, B.P. Flannery, S.A. Teukolsky, W.T. Vetterling: {\em Numerical
  Recipes in {FORTRAN}\/}, 2nd edn. (Cambridge University Press, Cambridge
  1992)

\bibitem{WanJL96}
Z.~Wang, B.~Jovanovic, D.H. Lee: Phys. Rev. Lett. {\bf 77\/}, 4426 (1996)

\bibitem{WeyJ98}
A.~Weymer, M.~Janssen: Ann. Phys. (Leipzig) {\bf 7\/}, 159 (1998).
  Cond-mat/9805063

\bibitem{AviBB99}
Y.~Avishai, Y.~Band, D.~Brown: Phys. Rev. B {\bf 60\/}, 8992 (1999)

\bibitem{CobK96}
D.H. Cobden, E.~Kogan: Phys. Rev. B {\bf 54\/}, R17\,316 (1996)

\bibitem{CaiRRS00} P.~Cain, R.A. {R\"{o}mer}, M.E. Raikh,
  M.~Schreiber: {\em Localization-Delocalization quantum Hall
    transition in the present of a quenched disorder} (2001).

\bibitem{AhoHU83}
A.~Aho, J.E. Hopcroft, J.D. Ullman: {\em Data Structures and Algorithms\/}
  (Addison-Wesley, New York 1983)

\bibitem{grassberger} P. Grassberger, (this volume).

\end{thebibliography}

\clearpage
\addcontentsline{toc}{section}{Index}
\flushbottom
\printindex

\end{document}